\documentclass{PoS}

\usepackage[utf8]{inputenc}
\usepackage{color}
\usepackage{graphicx}
\usepackage{bm}
\usepackage{bbm}

\usepackage{amsfonts}
\usepackage{amsmath}
\usepackage{amssymb}
\usepackage{amsthm}

\usepackage{psfrag}
\usepackage{pstricks,pst-grad,color,shadow}
\usepackage{dsfont}
\usepackage{tikz}
\usetikzlibrary{shapes,arrows,backgrounds}
\usepackage{slashed}

\definecolor{darkgreen}{rgb}{0,0.73,0}

\newcommand{\erw}[1]{\left \langle #1 \right \rangle}

\newcommand{\ii}{\mathrm{i}}
\newcommand{\id}{\mathbbm{1}}
\newcommand{\Nf}{N_\text{f}}
\newcommand{\Nc}{N_\text{c}}

\DeclareMathOperator{\tr}{Tr}

\graphicspath{{./}{../pdfPlots/}{../plots/}}

\title{$\mathcal{N}=1$ Supersymmetric $SU(3)$ Gauge Theory - Towards simulations of Super-QCD}

\ShortTitle{Simulations of Super-QCD}

\author{\speaker{Bj\"orn Wellegehausen}\\
        Friedrich Schiller University Jena, 07743 Jena, Germany\\
        E-mail: \email{bjoern.wellegehausen@uni-jena.de}}

\author{Andreas Wipf\\
       Friedrich Schiller University Jena, 07743 Jena, Germany\\
       E-mail: \email{wipf@tpi.uni-jena.de}}

\abstract{$\mathcal{N}=1$ supersymmetric QCD (SQCD) is a possible building block of theories beyond the standard model. It describes the interaction between gluons and quarks with their superpartners, gluinos and squarks. Since supersymmetry is explicitly broken by the lattice regularization, a careful fine-tuning of operators is necessary to obtain a supersymmetric continuum limit. For the pure gauge sector, $\mathcal{N}=1$ Supersymmetric Yang-Mills theory, supersymmetry is only broken by a non-vanishing gluino mass. If we add matter fields, this is no longer true and more operators in the scalar squark sector have to be considered for fine-tuning the theory. Guided by a one-loop calculation in lattice
	perturbation theory, we show that maintaining chiral symmetry in the light sector is nevertheless an important step. Furthermore, we present first preliminary lattice results on the fine-tuning and Ward-identities of SQCD.}

\FullConference{The 36th Annual International Symposium on Lattice Field Theory - LATTICE2018\\
		22-28 July, 2018\\
		Michigan State University, East Lansing, Michigan, USA.}

\begin{document}

\section{Introduction}
\noindent
Supersymmetric gauge theories are an important building block of many models beyond the standard model. The 
most simplest one is
$\mathcal{N}=1$ supersymmetric Yang-Mills theory (SYM), that describes the interaction of gluons with their superpartners, the gluinos. Supersymmetry is explicitly broken by the lattice formulation, but for this theory it has been shown that it is sufficient to restore chiral symmetry in the continuum limit to recover supersymmetry \cite{Curci:1986sm}. The mass spectrum and chiral properties of SYM theory have been extensively studied on the lattice with Wilson fermions for the gauge groups $SU(2)$ \cite{Bergner:2015adz} and $SU(3)$
 \cite{Ali:2018dnd,Steinhauser:2018}. The next step towards a supersymmetric standard model is the inclusion of matter fields in the fundamental representation, leading to supersymmetric QCD (SQCD). SQCD describes the interaction between gluons and quarks with their
  superpartners, gluinos and squarks. So far it has been studied with classical and semi-classical approaches \cite{Affleck:1983mk}, but non-perburbative results, especially for the mass spectrum of the theory, are
  not available. One reason is the large number of operators, that has to be fine-tuned in lattice simulations \cite{Giedt:2009yd}. A possible approach to
  control all relevant operators is a fine-tuning guided by lattice perturbation theory. This method has been successfully applied to two-dimensional supersymmetric gauge theories \cite{August:2018esp,Suzuki:2005dx}. In the present work we review the continuum formulation of SQCD and discuss our lattice formulation and fine-tuning of operators. Then we derive a one-loop effective potential for the squark field to improve our lattice action with respect to the restoration of supersymmetry in the continuum limit. Finally we present first lattice results regarding a possible sign problem of the theory and present our result for supersymmetric Ward-identities. A detailed discussion and further results will be published in a follow-up paper \cite{Wellegehausen:2018}.

\section{Continuum formulation of supersymmetric QCD}
\noindent
The Lagrange function of SQCD for gauge group $SU(N)$ with $\Nf$ flavours is given by
\begin{equation}
 \begin{aligned}
  \mathcal{L}=&\tr\left(-\frac{1}{4}F_{\mu\nu} F^{\mu\nu}+\frac{\ii}{2}\bar{\lambda} \gamma^\mu D_\mu \lambda\right)-D_\mu \phi^\dagger D^\mu \phi+\ii \bar{\psi} \gamma^\mu D_\mu \psi\\
  &-\ii \sqrt{2} g \left(\phi^\dagger \bar{\lambda} P \psi-\bar{\psi} \bar{P} \lambda \phi\right)-\frac{1}{2}g^2\left(\phi^\dagger T \sigma_3 \phi\right)^2+m\left(\bar{\psi}\psi-\frac{m}{Z} \phi^\dagger \phi\right) \label{eq:ContLagrange}
 \end{aligned}
\end{equation}
where the gluino field $\lambda$ is a Majorana spinor in the adjoint representation of the gauge group, 
the quark field $\psi$ a Dirac spinor and the squark field $\phi=(\phi_+,\phi_-)$ a two-component complex scalar field, both in the fundamental representation. The
components of
$P=(P_+,P_-)$ are the chiral projectors, 
$\bar{P}=\gamma_0 \, P\, \gamma_0$ and $T$ represents the generators of the gauge group $SU(N)$. The Pauli matrices $\sigma$ act on the two components of the squark field. The covariant derivatives are
\begin{equation}
 D_\mu (\psi\,,\phi)=\left(\partial_\mu +\ii g A_\mu \right) (\psi\,,\phi)\,, \quad \text{and} \quad D_\mu \lambda=\partial_\mu \lambda+\ii g [A_\mu,\lambda].
\end{equation}
The action is invariant under local $SU(N)$ gauge transformations as well as under supersymmetry transformations and a discrete parity transformation. The global symmetry group of the classical action is $SU(N_\text{f})_\text{L} \otimes SU(N_\text{f})_\text{R} \otimes U(1)_\text{A} \otimes U(1)_\text{B} \otimes U(1)_\text{R}$.
The $U(1)_\text{B}$ (baryon number conservation) is unbroken even in the quantum theory and restricts operators to vanishing baryon number. 
The chiral $U(1)_\text{A}$ symmetry and R-symmetry $U(1)_\text{R}$
\begin{equation}
  U(1)_\text{A}: \quad\psi \to e^{\ii \alpha \gamma_5} \psi\,, \quad \phi \to e^{\ii \alpha \sigma_3} \phi\,,\quad  U(1)_\text{R}: \quad \psi \to e^{-\ii \alpha \gamma_5} \psi\,, \quad \lambda \to e^{\ii \alpha \gamma_5}\lambda
\end{equation}
are broken in the quantum theory by the axial anomaly, similar to QCD and SYM theory. However, an anomalous free $U(1)_\text{AF}$ subgroup of $U(1)_\text{A} \otimes U(1)_\text{R}$ is still preserved in the quantum theory. 

\section{Lattice formulation and fine-tuning}
\noindent
The lattice formulation with Wilson fermions employed in this work breaks supersymmetry, chiral symmetry and R-symmetry explicitly. The remaining intact symmetries of the lattice theory restrict the number of operators that need to be fine-tuned in a lattice simulation to restore all symmetries in the continuum limit. A detailed discussion of the important operators and the lattice formulation can be found in \cite{Giedt:2009yd,Wellegehausen:2018}. The mass dimensions of the fundamental fields are $[A]=[\phi]=1$ and $[\psi]=[\lambda]=3/2$. Only relevant operators with mass dimension $[\mathcal{O}]\leq D=4$ have to be fine-tuned. Therefore, we have to consider only quadratic, cubic and quartic interactions in the following. To simplify the discussion, we consider only the gauge group $SU(3)$ with $N_\text{f}=1$. On a kinematic level the generalization to an arbitrary gauge group $SU(N)$ with fermion number $N_\text{f}$ is not difficult.

\paragraph*{Quadratic interactions}
Invariance under $SU(3)$ gauge transformations restricts fundamental scalar and fermionic bilinears to two different tensor products, namely
 $3 \otimes \bar{3}=1 \oplus 8$ and $8 \otimes 8=1 \oplus \dots$ .
The quarks and squarks transform in the fundamental representations $3$ and $\bar{3}$ of the gauge group while the gluino transforms under the adjoint representation $8$. Therefore, the only invariant combinations of component fields are
\begin{equation}
 \phi^\dagger \, \sigma \, \phi\,, \quad \bar{\psi}\,\Gamma\, \psi\quad \text{and} \quad \bar{\lambda}\, \Gamma\, \lambda \quad \text{with} \quad \Gamma=\{\id,\gamma_5\}\,,\quad \sigma=\{\id,\sigma_1\}.
\end{equation}
Quark and squark masses are compatible with supersymmetry and only the relative mass between quarks and squarks has to be tuned. An explicit gluino mass on the other hand breaks supersymmetry since its superpartner, the gluon, is always massless due to gauge invariance. The two possible scalar mass terms can be written as
\begin{equation}
 M_1=\tr \Phi \quad \text{and} \quad M_2=\tr (\Phi \, \sigma_1) \quad \text{with}\quad \Phi_{rs}=\phi^\dagger_{r} \phi_{s}\,,\quad r,s \in \{+,-\}.
\end{equation}
The second mass term $M_2$ breaks the axial chiral symmetry while the first is invariant under chiral transformations.

\paragraph*{Cubic interactions}
Invariance under gauge symmetry restricts the Yukawa interactions to the form
$ 3 \otimes 3 \otimes 3$ or $\bar{3} \otimes 8 \otimes 3.$
The first term is not invariant under the $U(1)_\text{B}$ symmetry and therefore, the only term invariant under all symmetries is the one already appearing in the Lagrangian \eqref{eq:ContLagrange}. Its renormalization can always be absorbed by a wave-function renormalization of the scalar fields. The Yukawa interaction 
should be a marginal operator in four dimensions.
 
\paragraph*{Quartic interactions}
The five marginal quartic operators, that have to be fine-tuned are
 \begin{equation}
 \begin{aligned}
  V_1=&\Phi_{++}^2+\Phi_{--}^2\,, \quad V_2=\Phi_{+-}^2+\Phi_{-+}^2\,,\quad V_3=\Phi_{++}\Phi_{--}\,,\\
  V_4=&\Phi_{+-}\Phi_{-+}\,,\quad V_5=\left(\Phi_{+-}+\Phi_{-+}\right)\left(\Phi_{++}+\Phi_{--}\right).
  \end{aligned}
\end{equation}
The operators $V_1$, $V_3$ and $V_4$ are invariant under the $U(1)_\text{A}$ transformations while $V_2$ and $V_5$ break chiral symmetry. The squark potential in the continuum action can be written as a linear combination of this five operators using $SU(N)$ Fierz identities.\\
After a rescaling of the fields and a Wick rotation we obtain the most general Euclidean Lagrange density:
\begin{equation}
\begin{aligned}
  \mathcal{L}=&\frac{1}{g^2}\left(\frac{1}{4} \tr F_{\mu\nu}^2+Z_{\phi}\, D_\mu \phi^\dagger D_\mu \phi+Z_{\phi} \,m_i^2\, M_i+Z_{\phi}^2\,\lambda_i\, V_i\right)\\
  &+\frac{1}{2} \tr \bar{\lambda} \left( \gamma_\mu D_\mu -m_\text{g}\right) \lambda + \bar{\psi}\left(\gamma_\mu D_\mu -m_\text{q}\right)\psi+\ii \sqrt{2} \left(\phi^\dagger \bar{\lambda} \, P \, \psi-\bar{\psi} \, \bar{P} \, \lambda \, \phi\right).
 \end{aligned}
\end{equation}
The relation to the continuum action \eqref{eq:ContLagrange} is given by
\begin{equation}
m_\text{g}=0\,,\;m_\text{q}=m_1=m\,,\;Z_\phi=1\,,\;\lambda_1=\frac{\Nc-1}{\Nc}\,,\;\lambda_3=\frac{1}{\Nc}\,,\;\lambda_4=-1\,,\;m_2=\lambda_2=\lambda_5=0. \label{eq:contParam}
\end{equation}
To restore supersymmetry in the continuum limit of the corresponding lattice action, we have to fine-tune $9$ operators for $N_\text{f}=1$. This are the bare gluino mass $m_\text{g}$, the two squark masses $m_i$, the squark wave function renormalization $Z_\phi$ and the five squark quartic couplings $\lambda_i$. To reduce the number of operators, we will fix as many as possible by a perturbative calculation of the lattice squark effective potential in our first simulations of SQCD.

\section{One-loop effective potential}
\noindent
Very recently, SQCD has been investigated with lattice perturbation theory and critical fermion and squark masses have been computed to one-loop order \cite{Costa:2017rht}. Here, we compute an effective squark potential at one-loop in lattice perturbation theory. It contains a bosonic and a fermionic contribution, that come with a different sign
\begin{equation}
V_\text{eff}^{\text{1-loop}}(\phi)=V(\phi)+\frac{1}{2}\tr \ln\left(\left.\frac{\delta^2 S[\varphi]}{\delta \varphi(x)\delta \varphi(y)}\right\vert_{\varphi=\phi}\right)-\frac{1}{2}\tr\ln\left(\frac{\delta^2 S[\phi]}{\delta \Psi(x)\delta \Psi(y)}\right).
\end{equation}
Before momentum integration we obtain the one-loop correction
\begin{equation}
V(p)=C_\text{F}\left(2\Delta_{\phi}+6 \Delta _G-8 \Delta_\phi\right)\,M_1 \\
+f_1(\Delta_{\phi}^2,\Delta_G^2)\,V_1+f_3(\Delta_{\phi}^2,\Delta_G^2)\,V_3+f_4(\Delta_{\phi}^2,\Delta_G^2)\,V_4
\end{equation}
with scalar field propagator $\Delta_\phi=1/(p^2+m^2)$ and gluon propagator $\Delta_G=1/p^2$. $C_\text{F}$ is the quadratic Casimir eigenvalue of $SU(\Nc)$ and the functions $f_i$ depend on the squark and gluon propagator, the number of colors $\Nc$ and the quark / squark mass $m$. We observe, that at one-loop in perturbation theory, only the operators $M_1$, $V_1$, $V_3$ and $V_4$, that are already present in the tree-level action, appear. Furthermore, the quadratic divergences cancel and the functions $f_i$ contain only logarithmic divergences as expected from supersymmetry.\\
On the lattice, the Wilson mass and discrete lattice momenta break supersymmetry. We have to replace the continuum propagators by the corresponding lattice propagators. As a consequence, all possible $7$ squark operators are generated at one-loop, and we have to introduce counterterms, such that the lattice action at one-loop corresponds to the continuum action at one-loop. The values of the couplings $c_i$ for the $7$ counterterms are given in Table~\ref{tab:Counter0} for $m=0$ and different lattice volumes.
\begin{table}[htb]
\begin{tabular}{|c||c|c|c|c|c|c|c|}
\hline
$L$ & $c_1 (M_1)$ & $c_2 (M_2)$& $c_3 (V_1)$ & $c_4 (V_2)$ & $c_5 (V_3)$ & $c_6 (V_4)$ & $c_7 (V_5)$\\
\hline\hline
 8 &  -1.379358 & -0.6361834 & 0.3057303 & -0.05689831 & 0.252934 & 0.21088 & 0.077108 \\
 16 & -1.378523 & -0.6366576 & 0.3101264 & -0.05722842 & 0.258885 & 0.22182 & 0.083166 \\
 32 & -1.378457 & -0.6366606 & 0.3108112 & -0.05723039 & 0.259836 & 0.22434 & 0.084368 \\
 64 & -1.378453 & -0.6366606 & 0.3109568 & -0.05723041 & 0.260038 & 0.22496 & 0.084646 \\
 128 & -1.378453 & -0.6366606 & 0.3109917 & -0.05723041 & 0.260086 & 0.22511 & 0.084714 \\
 256 & -1.378453 & -0.6366606 & 0.3110003 & -0.05723041 & 0.260098 & 0.22515 & 0.084731 \\
 \hline
 $\infty$ & -1.378453 & -0.6366606 & 0.3110030 & -0.05723041 & 0.260102 & 0.22516 & 0.084736 \\
 \hline
\end{tabular}\hskip10mm
\caption{Couplings $c_i$ of the counterterms obtained from the one-loop effective potential for $m=0$ and $\Nc=3$ and different lattive volumes $V=L^3 \times 2 L$.\label{tab:Counter0}}
\end{table}
It turns out, that the counterterms are finite and their infinite volume extrapolation is given in the last line of the table. In our simulations we use the values at the given lattice size and squark mass.
In the following, the unimproved lattice action corresponds to the parameters given in \eqref{eq:contParam} while the improved action is given by free parameters $m_\text{g}$, $m_\text{q}$ and
\begin{equation}
m_1^2=m^2+g^2\,c_1\,, \quad m_2^2=g^2\,c_2\,, \quad \lambda_i=\lambda_{i,c}+g^2\,c_{i+2} \quad \text{with} \quad \lambda_{i,c} \quad \text{given by \eqref{eq:contParam}}.
\end{equation}

\section{Lattice results}
\begin{figure}[htb]
\hskip5mm\scalebox{0.92}{\includegraphics{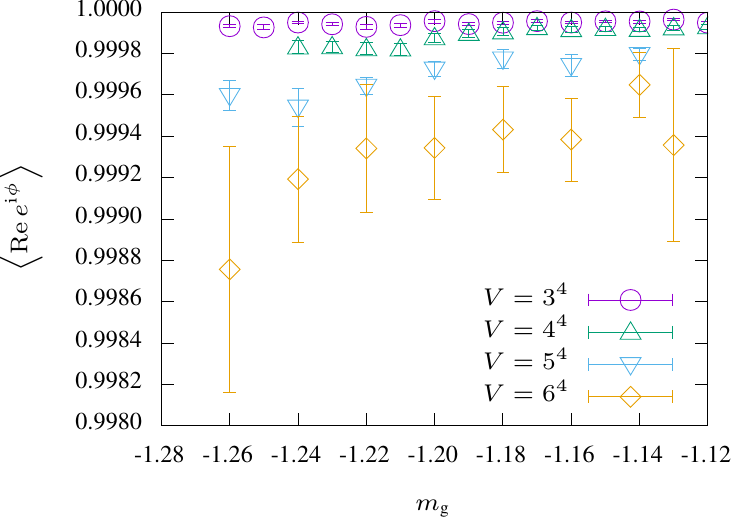}}\hskip5mm
\scalebox{0.92}{\includegraphics{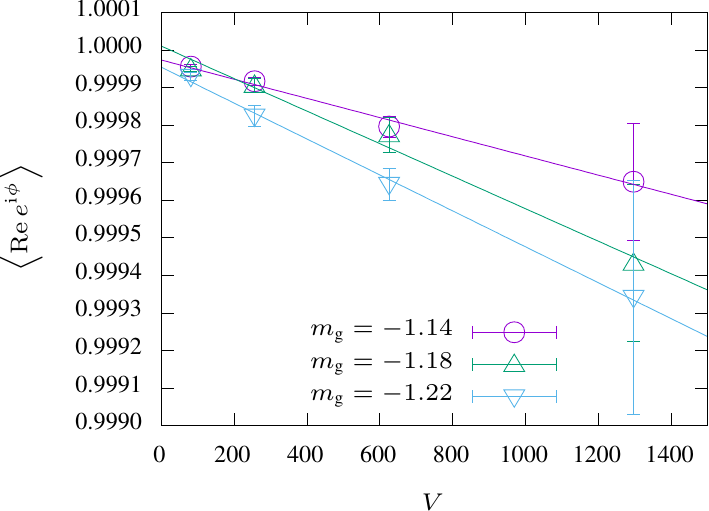}}
\caption{Pfaffian phase as function of the bare gluino mass $m_\text{g}$ for different lattice volumes (left panel) and as function of the lattice volume for three different bare gluino masses (right panel).\label{Fig:sign}}
\end{figure}

\begin{figure}[tb]
\hskip10mm\scalebox{0.92}{\includegraphics{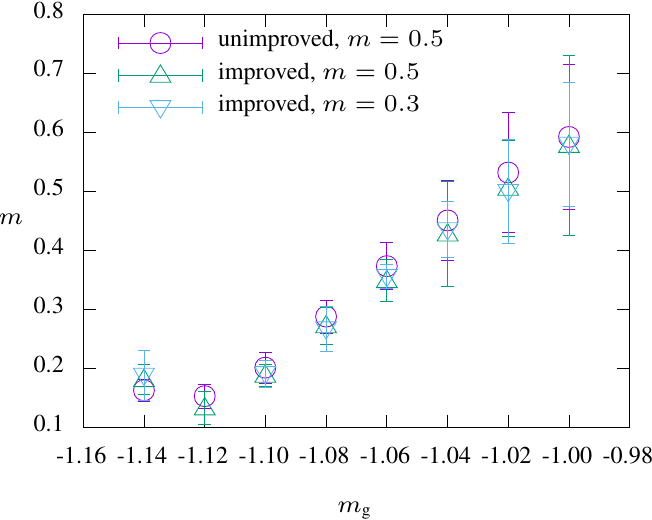}}\hskip5mm
\scalebox{0.92}{\includegraphics{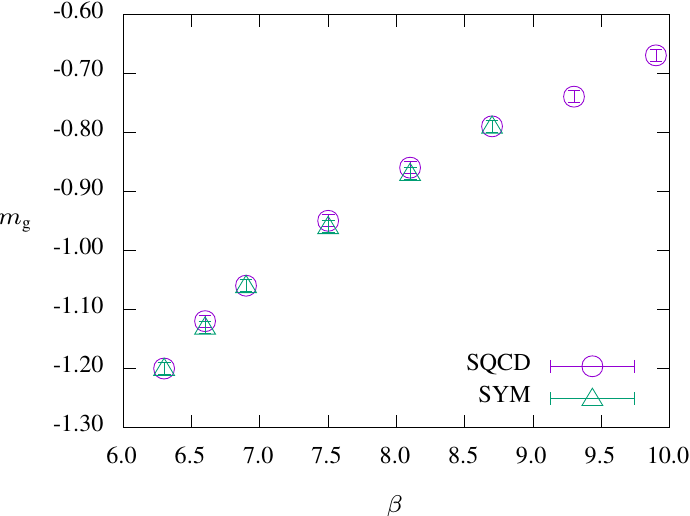}}
\caption{Adjoint pion mass (left panel) for $\beta=6.6$ and critical line (right panel) on a $8^3 \times 16$ lattice as function of the inverse gauge coupling $\beta$.\label{Fig:adjointPion}}
\end{figure}

\begin{figure}[tb]
\scalebox{0.71}{\includegraphics{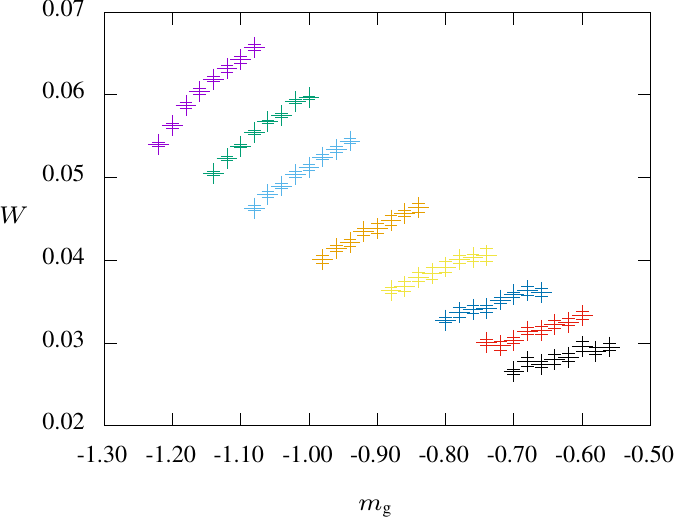}}
\scalebox{0.71}{\includegraphics{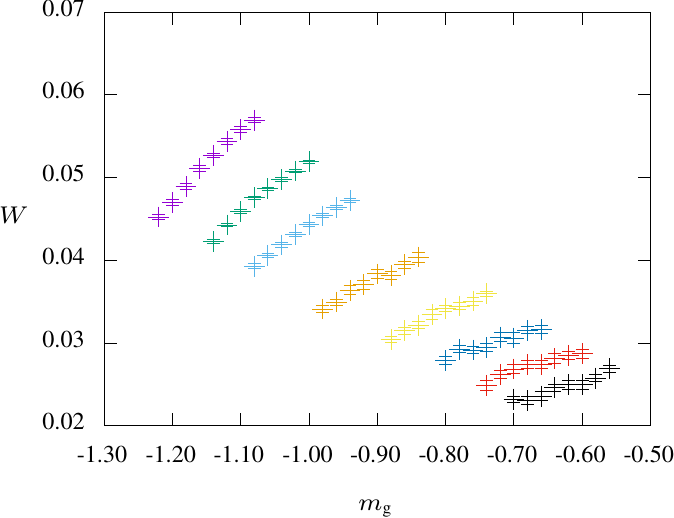}}
\scalebox{0.71}{\includegraphics{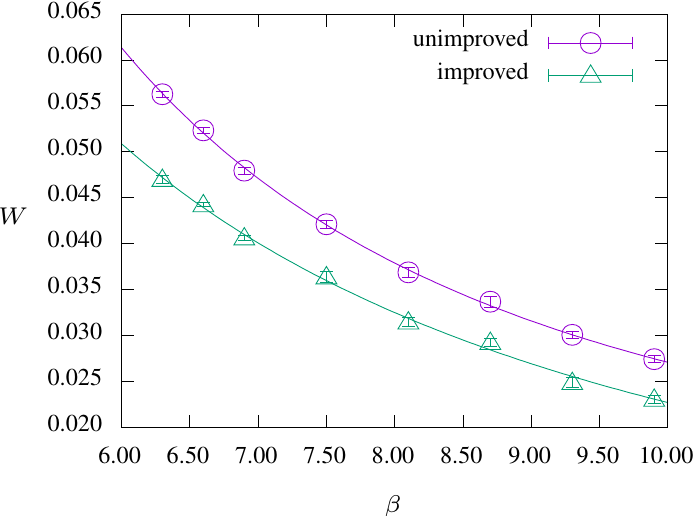}}
\caption{Bosonic Ward-identity $W$ for the unimproved (left panel) and improved (center panel) action as function of the gluino mass. The different colors encode different (from left to right increasing) values of $\beta$. In the right panel, the bosonic Ward-identity at the critical gluino mass $m_\text{g}^\text{c}$ for the unimproved and improved action is shown. The solid lines represent fits to the data points.\label{Fig:Ward}}
\end{figure}

\noindent
In a first test of our fine-tuning procedure 
we consider the limit of heavy quarks and squarks, where
 we expect the theory to behave similar to $\mathcal{N}=1$ SYM theory. The simulations are performed on lattice volumes up to $8^3 \times 16$ with inverse gauge couplings $\beta=\Nc/g^2=6.3$ to $\beta=9.9$ at $m=0.3$, $m=0.5$ and $m_\text{q}=-0.4$. In this regime, we expect that the explicit breaking of chiral symmetry dominates the spontaneous breaking. First we investigate the Pfaffian sign problem induced by the Yukawa interaction between gluinos, quarks and squarks. On rather small lattices we compute the Pfaffian exactly as function of the bare gluino mass $m_\text{g}$. The result for the Pfaffian phase expectation value $\erw{\text{Re}\,\exp\{\ii\phi\}}$ is shown in the left panel of Figure~\ref{Fig:sign}.
The sign problem increases with the lattice volume and depends only slightly on the distance to the critical point. On these very small lattices where we can compute the exact Pfaffian, the phase is very close to one. In the right panel we show the sign as function of the lattice volume for three different gluino masses. The solid lines are linear fits to the data points. Extrapolated to larger volumes we obtain a phase expectation value of approximately $0.996 - 0.998$ at $V=8^3 \times 16$ and $0.94 - 0.97$ at $V=16^3 \times 32$. Even at this large lattice the sign problem is almost absent and we can simulate the theory without taking into account the Pfaffian phase.\\
Similar to $\mathcal{N}=1$ SYM theory we determine the parameters for vanishing gluino mass by fine-tuning the theory to vanishing adjoint pion mass. The adjoint pion is not a physical particle but its correlation function is calculated in analogy to QCD as the connected part of the adjoint $\eta$-meson correlation function. The extracted adjoint pion mass is shown in the left panel of Figure~\ref{Fig:adjointPion} on a $8^3 \times 16$ lattice as function of the bare gluino mass $m_\text{g}$ and inverse gauge coupling $\beta=6.6$.
The results is rather insensitive to the 
choice of mass parameter $m=0.5$ or $m=0.3$ and the improved or unimproved action, indicating that indeed in this limit the light gluino sector and the heavy quark / squark sector decouple.
We obtain a critical bare gluino mass of $m_\text{g}^\text{c}=-1.12(1)$. In the right panel we show the critical mass $m_\text{g}^\text{c}$ as function of $\beta$ compared to the critical line of $\mathcal{N}=1$ SYM theory. As expected, the critical lines agree in the limit of heavy quarks and squarks.\\
Next we investigate, how well the the bosonic Ward-identity
\begin{equation}
W=\frac{W_\text{c}}{14}-1=0\,,\quad W_\text{c}=\frac{2}{3}\erw{\!-\frac{1}{4}\tr F_{\mu\nu}F^{\mu\nu}}+\erw{ D_\mu\phi^\dagger D^\mu \phi+m^2 \phi^\dagger \phi}=\Nc^2-1+2 \Nc \Nf
\end{equation}
is fulfilled
for the unimproved and improved actions on the $8^3 \times 16$ lattice. In Figure~\ref{Fig:Ward} $W$ is shown for the unimproved (left panel) and improved (center panel) action as function of the inverse gauge coupling and bare gluino mass.
The breaking of the Ward-identity increases with increasing $m_\text{g}$ while it decreases with 
increasing $\beta$. At fixed values of $m_\text{g}$ and $\beta$, it is less broken for the improved action. We then fit the Ward-identity $W(m_\text{g}^\text{c},\beta)$ at the critical gluino mass to the function
\begin{equation}
W(\beta)=W_\infty+a\, \beta^\kappa.
\end{equation}
The result is shown in Figure~\ref{Fig:Ward} (right panel). We obtain in the limit $\beta \to \infty$ the fit values $W_\infty^\text{unimproved}=0.008(3)$ and $W_\infty^\text{improved}=-0.003(11)$. Both results are compatible with a restored bosonic Ward-identity in this limit. At finite lattice spacing, the Ward-identity 
$W$ with improved action is always closer to its continuum value $0$ than for the unimproved action.

\section{Conclusions}
\noindent
Compared to $\mathcal{N}=1$ SYM theory on the lattice, supersymmetric QCD requires much more operators to be fine-tuned to perform a supersymmetric continuum limit. In this work we discussed the relevant operators in a lattice formulation with Wilson fermions and calculated a one-loop potential for the squark field to guide the fine-tuning towards a continuum limit with restored supersymmetry. We simulated supersymmetric QCD in the limit of heavy quarks and squarks and showed, that the fermion sign problem induced by the Yukawa interaction is almost absent. Furthermore, our one-loop improved lattice action leads to an improved bosonic Ward-identity that seems to be fulfilled in the continuum limit. So far, simulations have been done for heavy quarks and squarks, and the theory behaves similar to SYM theory. Simulations for lighter matter fields and a detailed discussion of our lattice formulation and the calculation of the one-loop effective squark potential will be made public in \cite{Wellegehausen:2018}.
\\
\textbf{Acknowledgments:}
We thank Stefano Piemonte, Marios Costa, Haralambos Panagopoulos, Andre Sternbeck and Marc Steinhauser for discussions and helpful comments.

\bibliographystyle{JHEP}
\bibliography{references}

\end{document}